\documentclass[twocolumn]{aastex61}
\usepackage{amsmath}
\usepackage{amsmath,amstext}
\usepackage[T1]{fontenc}
\usepackage{color}
\usepackage{cleveref}

\listofchanges
\begin{document}
\title{Revisiting the Energy Budget of WASP-43b: Enhanced day-night heat transport}

\author{Dylan Keating}
\affiliation{Department of Physics, McGill University, 3600 rue University, Montr\'eal, QC, H3A 2T8, CAN}
\author{Nicolas B. Cowan}
\affiliation{Department of Physics, McGill University, 3600 rue University, Montr\'eal, QC, H3A 2T8, CAN}
\affiliation{Department of Earth \& Planetary Sciences, McGill University, 3450 rue University, Montreal, QC, H3A 0E8, CAN}

\begin{abstract}
The large day--night temperature contrast of WASP-43b \citep{2014Sci...346..838S,2017AJ....153...68S} has so far eluded explanation \citep[e.g.,][]{2015ApJ...801...86K}. We revisit the energy budget of this planet by considering the impact of reflected light on dayside measurements, and the physicality of implied nightside temperatures. Previous analyses of the infrared eclipses of WASP-43b have assumed reflected light from the planet is negligible and can be ignored. We develop a phenomenological eclipse model including reflected light and thermal emission and use it to fit published \emph{Hubble} and \emph{Spitzer} eclipse data. We infer a near-infrared geometric albedo of 24$\pm$1\% and a cooler dayside temperature of $1483 \pm 10~$K. Additionally, we perform lightcurve inversion on the three published orbital phase curves of WASP-43b and find that each suggests unphysical, negative flux on the nightside.  By requiring non-negative brightnesses at all longitudes, we correct the unphysical parts of the maps and obtain a much hotter nightside effective temperature of $1076 \pm 11~$K. The cooler dayside and hotter nightside suggests a heat recirculation efficiency of 51\% for WASP-43b, essentially the same as for HD 209458b, another hot Jupiter with nearly the same temperature. Our analysis therefore reaffirms the trend that planets with lower irradiation temperatures have more efficient day\--night heat transport. Moreover, we note that 1) reflected light may be significant for many near-IR eclipse measurements of hot Jupiters, and 2) phase curves should be fit with physically possible longitudinal brightness profiles --- it is insufficient to only require that the disk-integrated lightcurve be non-negative. 
\end{abstract}

\section{Introduction} \label{sec:intro}
In many ways, the hot Jupiter WASP-43b \citep{2011A&A...535L...7H} is like other planets of this classification. It has a radius of $1.036\pm0.019~R_{\rm J}$, a mass of $2.034\pm0.052~M_{\rm J}$, and an orbital period of $0.81$ days \citep{2012A&A...542A...4G}. However, unlike other hot Jupiters, it orbits a fairly cool K7V star. Secondary eclipses of WASP-43b have been observed in multiple photometric bands \citep{2013ApJ...770...70W,2014ApJ...781..116B,2014A&A...563A..40C,2014MNRAS.445.2746Z}. Full orbit phase curves from Hubble Space Telescope Wide Field Camera 3 (WFC3) using the G141 grism (1.1--1.7$~\mu$m), and \emph{Spitzer} Infrared Array Camera (IRAC) at $3.6$ and $4.5~\mu$m, were used to retrieve phase resolved emission spectra \citep{2014Sci...346..838S,2017AJ....153...68S}. Emission and transmission spectroscopy measurements with WFC3 were used by \citet{2014ApJ...793L..27K} to determine the precise amount 
of water in the atmosphere of WASP-43b. The planet's transit times are consistent with a constant period and show no evidence of orbital decay \citep{2016AJ....151..137H}.

Previous analyses of WASP-43b reported an eastward hotspot offset that is typical of hot Jupiters and almost nonexistent heat transport from dayside to nightside \citep{2014Sci...346..838S,2015MNRAS.449.4192S,2017arXiv170705790S}. The three-dimensional atmospheric circulation models of \citet{2015ApJ...801...86K} were able to provide a good match to the WFC3 dayside emission spectrum and were able to reproduce the eastward offset by invoking equatorial superrotation. The model nightside, however, was too bright (hot) compared to the low measured nightside flux. Nightside clouds have been postulated as a way to explain the low measured nightside flux \citep{2015ApJ...801...86K,2017AJ....153...68S}. Clouds or not, if the observations are taken at face value, then WASP-43b does a poor job of moving heat from day to night.  This is in stark contrast to theoretical expectations and empirical trends, both of which favor increasing day--night temperature contrast with increasing irradiation \citep{2011ApJ...729...54C,2013ApJ...776..134P,2015MNRAS.449.4192S,Komacek2016}.  

All analyses of HST/WFC3 1.1--1.7$~\mu$m exoplanet secondary eclipses have assumed that the reflected light component in this bandpass is negligible compared to the thermal emission component. \citet{2007ApJ...667L.191L} showed that thermal emission dominates reflected light for highly irradiated hot Jupiters with low Bond albedo and inefficient heat redistribution. Previous studies have found that most hot Jupiters have very low geometric albedos at optical wavelengths \citep{2008ApJ...689.1345R,2011MNRAS.417L..88K,2013ApJ...777..100H,2017AJ....153...40D} and it has since been taken for granted that reflected light is also negligible in the near infrared. However, \citet{2007ApJ...667L.191L} also showed that for planets with efficient heat redistribution and Bond albedo of 50\%, reflected light can instead dominate thermal emission in the near infrared. \citet{2015MNRAS.449.4192S} found a systematic offset between Bond albedos inferred from thermal phase variations and geometric albedos obtained from visible light measurements; they suggested that hot Jupiters may reflect 30--50\% of incident near-infrared radiation. 

If a hot Jupiter reflects light at a given wavelength, then the eclipse depth will be greater, and one will infer too high a dayside temperature if the reflected light is ignored. Since total bolometric flux is proportional to the fourth power of temperature, even a small change in temperature leads to a significant change in bolometric flux. 
 
With the notable exception of the earliest high-cadence phase curve measurements \citep{2007Natur.447..183K, 2009ApJ...690..822K}, exoplanet researchers have been content to publish phase curve parameters without worrying about the particular brightness distribution that could give rise to such a lightcurve \citep{2008ApJ...678L.129C}.  Instead, theorists have taken the extra step of producing mock observations, which can be compared to the real thing.  In the few cases where theory and the observations have not matched up, it has been attributed to missing physics in the models, rather than unphysical phase curves. 

In Section~2 of this letter, we revisit the dayside measurements, accounting for reflected light to obtain a new dayside effective temperature for WASP-43b. In Section~3, we revisit the nightside measurements, correcting for negative brightnesses at certain longitudes, to obtain a new nightside effective temperature. In Section~4, we use these two new effective temperatures to re-estimate the Bond albedo and heat recirculation efficiency for WASP-43b, and we discuss the implications for that planet as well as other hot Jupiters.

\section{Thermal Plus Reflected Eclipse Model}\label{a}
Immediately before and after a secondary eclipse, the flux we observe from a planet is some combination of reflected starlight and thermal emission \begin{equation}\label{eq:1}
 \frac{F_{\rm day}}{F_{*}} = A_{g} \left( \frac{R_{p}}{a} \right) ^2+ \frac{B_{\lambda}(T_{\rm day})}{B_{\lambda}(T_{*})}
 \left( \frac{R_{p}}{R_{*}} \right)^2 
 \end{equation} 
\citep{2007MNRAS.379..641C}.
The eclipse depth, $F_{\rm day}/{F_{*}}$, is the ratio of the planet's dayside flux to the stellar flux at a particular wavelength. The star's brightness temperature at this wavelength is $T_{*}$, and $T_{\rm day}$ is the planet's dayside brightness temperature \added{at that wavelength.} The geometric albedo, $A_{g}$, is the fraction of starlight that the planet reflects back toward the star (and hence the observer), \added{and is also wavelength dependent.} The planetary and stellar radii are given by $R_{p}$ and $R_{*}$ respectively, and $a$ is the semimajor axis of the planet's orbit.

\begin{figure}[htb]
\epsscale{1.2}
\plotone{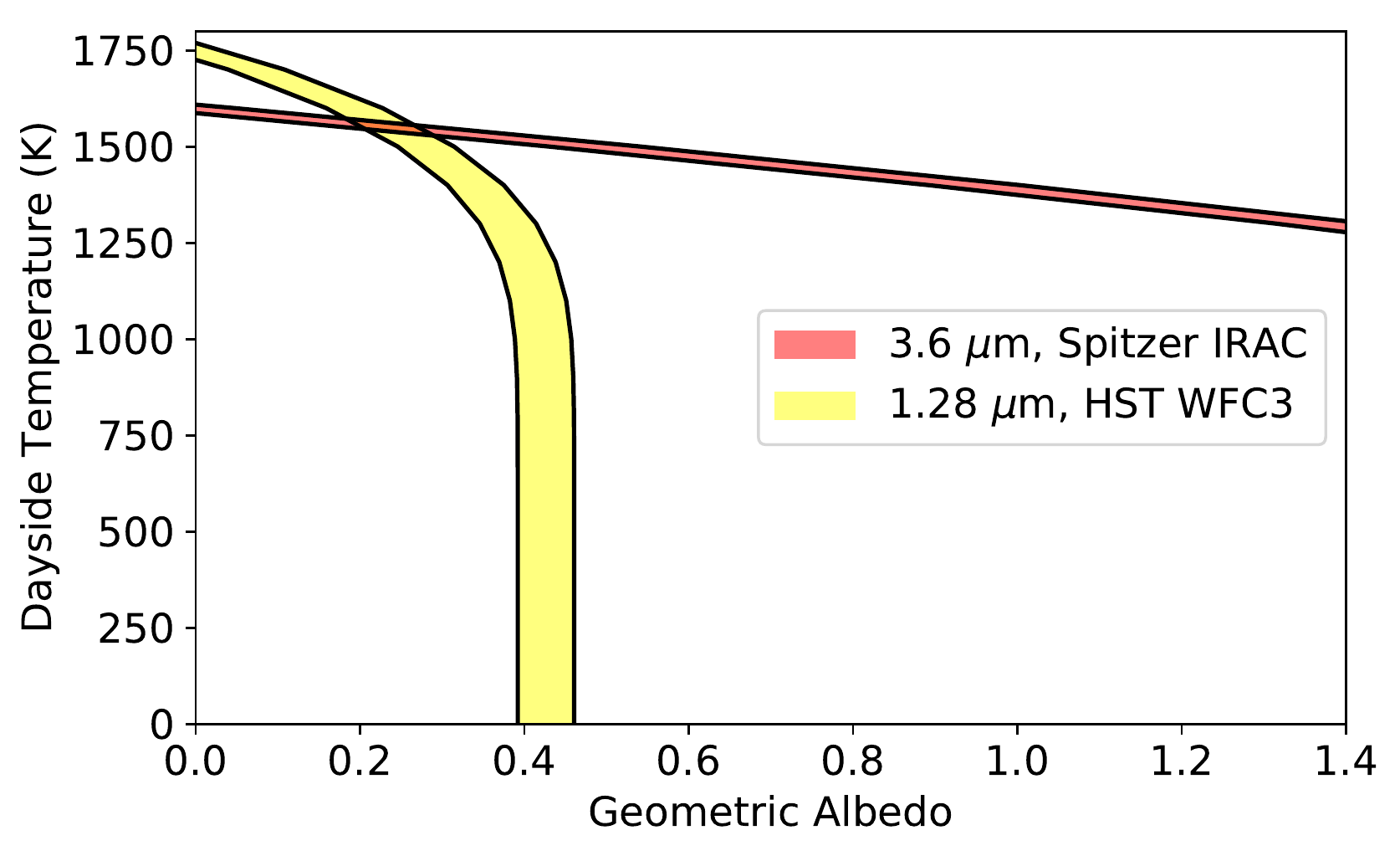}
\caption{Degeneracy between thermal emission and reflected light for WASP-43b. The 1$\sigma$ constraints from two published eclipse depths are shown. When $A_{g}=0$, only thermal emission contributes to the eclipse depth. As $A_{g}$ is increased, the amount of thermal emission decreases, and consequently so does $T_{\rm day}$.  Only two wavelengths are shown here for clarity, but the following trend holds: the \emph{Spitzer} IRAC eclipse depth lines never touch the horizontal axis for physically possible values of geometric albedo, meaning \emph{Spitzer} measurements require thermal emission regardless of the value of $A_{g}$. For the HST/WFC3 measurements, the eclipse depths can be attributed solely to reflected light if the geometric albedo is $\sim40\%$. It may not be safe to ignore reflected light in the HST/WFC3 1.1--1.7$~\mu$m bandpass. \label{bananas}}
\end{figure}

Given an eclipse depth, the geometric albedo and dayside brightness temperature are inversely related, as can be seen in Figure \ref{bananas}. For the \emph{Spitzer}/IRAC $3.6$ and $4.5~\mu$m channels, thermal emission dominates and the reflected light term can be ignored. Even a geometric albedo in excess of unity can only account for a small fraction of the measured flux.  For the HST/WFC3 1.1--1.7$~\mu$m wavelengths, the reflected light component is usually also neglected. However as Figure \ref{bananas} shows, the WFC3 eclipse depths can be attributed solely to reflected light for even modest values of geometric albedo --- in this case $\sim 40\%$, which is consistent with estimates of Bond albedos for hot Jupiters \citep{2015MNRAS.449.4192S}, and theoretical predictions of geometric albedo for very hot planets \citep{2000ApJ...538..885S}.  

\added{In general, we expect different brightness temperatures at different wavelengths as they should probe different depths in the atmosphere. However, Figure 1 of \citet{2015MNRAS.449.4192S} shows that the aggregate broadband brightness temperature spectrum of 50 hot Jupiters is flat and featureless. They attributed this to a vertically isothermal atmosphere, optically thick clouds, or both. Even without clouds, the band-integrated infrared dayside brightness temperatures of hot Jupiters are predicted to be within $\sim$100~K of the dayside effective temperature; using brightness temperatures in three broadbands to estimate the dayside effective temperatures for these planets should only introduce a systematic error of 4--5\% \citep{2011ApJ...729...54C}.  For our analysis we treat the dayside atmosphere of WASP-43b as isothermal and fit its emission with a blackbody.} 

We use the model of reflected light plus thermal emission from \cref{eq:1} to fit the published secondary eclipse depths of WASP-43b from HST/WFC3 and \emph{Spitzer}/IRAC  \citep{2014Sci...346..838S,2017AJ....153...68S}; schematically, this is simply where the swaths intersect in Figure~\ref{bananas}, \added{but incorporating all of the eclipse depths.} \added{We used a Phoenix stellar model for the spectrum of the host star \citep{2000ApJ...539..366A}}. A gray reflectance was assumed, meaning a constant albedo for all wavelengths (in practice this should be taken to be the albedo in the WFC3 bandpass). We follow the lead of \citet{2017AJ....153...68S} and fit only the WFC3 and \emph{Spitzer} data. Unsurprisingly, our model is also a bad fit to the ground-based photometric data (see \Cref{fits}). \added{We omit data in the water band (\mbox{1.35--1.6$~\mu$m}) from our fit.\footnote{We get a good match to data in the H$_{2}$O band between \mbox{1.35--1.6$~\mu$m} by fitting the characteristic water feature from \citet{2016ApJ...823..109I}. However, the fitting is completely empirical and provides no additional information about the atmosphere of WASP-43b.}} Using a Markov Chain Monte Carlo \citep{2013PASP..125..306F}, we fit for the planet's geometric albedo and the dayside temperature. We find $A_{g} =0.24\pm0.01 $ and $T_{\rm day} = 1483\pm10$~K. A thermal-only model, with dayside temperature as the only parameter, yields T$_{\rm day}=1575 \pm7$~K. The Bayesian Information Criterion \citep[BIC;][]{schwarz1978} is much lower for the reflected plus thermal model than the thermal-only model ($\Delta$ BIC = 277), meaning we can strongly reject the thermal-only model in favor of the model with reflected light. Our fits are summarized in Table~\ref{fits}. 

Omitting the water band data means we are unable to directly compare $\Delta$ BIC between our toy model and the 6-parameter spectral retrieval of \citet{2017AJ....153...68S}. A full atmospheric retrieval, with the addition of reflected light, may be necessary for a statistically and physically complete model of WASP-43b's dayside. A comprehensive understanding of the dayside of WASP-43b should also address why the ground-based data disagree with the models. 

\floattable
\begin{deluxetable*}{ccccccccc}[htb!]
\tabletypesize{\small}
\tablecaption{Fit statistics for different combinations of eclipse depth data. We adopt the parameters from the fit that omits the water band, and incorporates only WFC3 and \emph{Spitzer} data, in order to be consistent with \citet{2017AJ....153...68S}. 
\label{fits}}
\tablenum{1}
\tablehead{\colhead{Data Used} & \colhead{Model} & \colhead{$T_{\rm day}$ (K)} & \colhead{$A_{g}$} & \colhead{$N_{\rm params}$} & \colhead{$N_{\rm data}$} & \colhead{$\chi^{2}$}&\colhead{$\chi^{2}$/Datum} & \colhead{BIC} 
} 
\startdata
WFC3, \emph{Spitzer}, no water band &  {\bf Thermal + Reflected}& $1483\pm10$ & $0.24\pm0.01$ & 2 & 9 & \textbf{53 }&5.9 & {\bf 58} \\
WFC3, \emph{Spitzer}, no water band &Thermal Only& $1575\pm 7$& & 1 & 9 & 333& 37 &  335 \\
\hline
All, no water band &  {\bf Thermal + Reflected} & $1331\pm 482$ & $0.36\pm0.18$ & 2 & 22 & {\bf 332}&20 & {\bf 337} \\
All, no water band &  Thermal Only & $1728\pm 163$ & & 1 & 22 & 614 &47& 616 \\
\enddata
\end{deluxetable*}

\begin{figure}[htb]
\epsscale{1.2}
\plotone{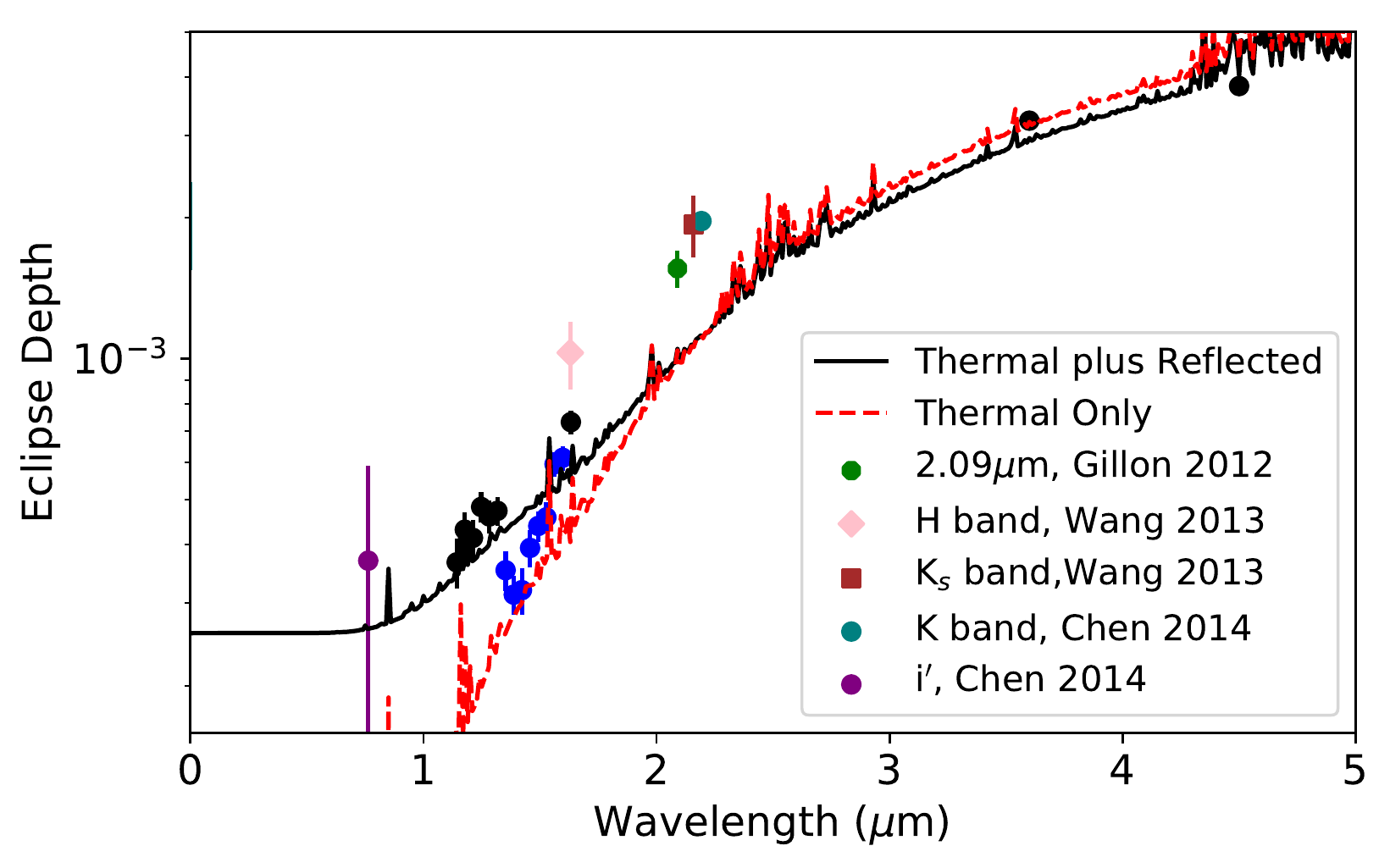}
\caption{WFC3 and \emph{Spitzer} IRAC eclipse depths fit with the two different toy models. The black points are the WFC3 and \emph{Spitzer} points. The water band points (in blue) are omitted from the fits. Photometric eclipse depths in different bands are shown but are also not incorporated in the fits, following \citep{2014Sci...346..838S,2017AJ....153...68S}.  The reflected light plus thermal emission model is preferred over the thermal-only model ($\Delta BIC = 277$). \label{spectrum}}
\end{figure}

\section{Revisiting the phase variations of WASP-43b}
\label{b}
\begin{figure}[htb]
\epsscale{1.2}
\plotone{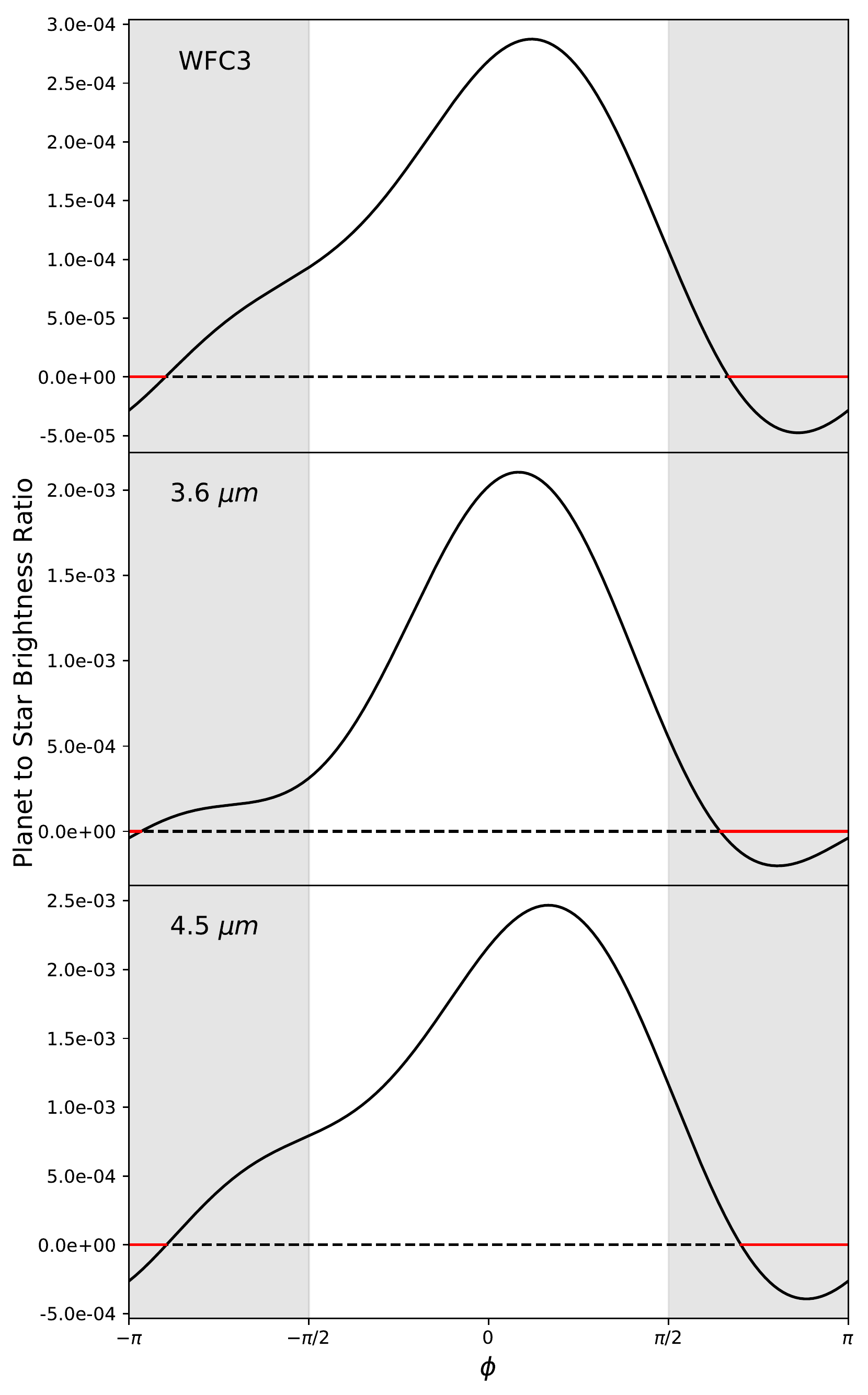}
\caption{Brightness maps corresponding to the three full orbit phase curves of WASP-43b \citep{2014Sci...346..838S,2017AJ....153...68S}. Here $\phi$ is the longitude from the substellar point and the nightside of the planet is shown as the grey shaded area. All three phase curves require negative brightnesses at certain longitudes, which is unphysical.  The red line shows where we doctor the maps by setting the brightness to zero.
\label{maps}}
\end{figure}

Since WASP-43b is on a circular, edge-on orbit and expected to be tidally locked, we can use Equation~7 from \cite{2008ApJ...678L.129C} to invert the phase curves, $F(\xi)$, into longitudinally resolved brightness maps, $J(\phi)$, where $\xi$ is the planet's phase angle ($\xi = 0$ at secondary eclipse, $\xi = \pi$ at transit), and $\phi$ is longitude from the substellar point (where $-\pi/2 < \phi < \pi/2$ is the dayside of the planet). \added{Since the phase curves were each fit using the fundamental frequency and its first harmonic (one and two cycles per orbit, respectively), the corresponding brightness maps also have two sinusoidal frequencies. \added{Higher frequencies in the map are assumed to be zero, following \citet{2008ApJ...678L.129C}.}} As can be seen in Figure~\ref{maps}, all three published phase curves require certain longitudes on the nightside of WASP-43b to have negative brightness, which is physically impossible. To properly estimate the nightside temperature of this planet, we require a map with non-negative brightness values at all longitudes. 

For each brightness map, $J(\phi)$, we keep the map as-is but set the negative brightnesses to zero. We then compute the phase curve for this doctored map using \citep{2008ApJ...678L.129C}:
\begin{equation}\label{eq:3}
F(\xi)=\int_{-\xi-\frac{\pi}{2}}^{-\xi+\frac{\pi}{2}}J(\phi)\rm{cos}(\phi+\xi)d\phi.
\end{equation}   
Evaluating this phase curve at $\xi=\pi$ gives the \added{ratio of disk-integrated} nightside flux \added{to stellar flux,} $F_{\rm night}$. We then calculate a nightside brightness temperature at each wavelength using 
\begin{equation}
T_{b}(\lambda) =\frac{hc}{\lambda k}\left[ \added{\ln} \left(1+\frac{e^{hc/\lambda k T_{*}}-1}{F_{\rm night}/\delta_{\rm tra}}\right)\right]^{-1} ,
\end{equation} 
where $\delta_{\rm tra} \added{= (R_{p}/R_{\star})^{2}}$ is the transit depth. For the brightness temperature of the star at a given wavelength, $T_{\star}$, we use Phoenix stellar grid models. Applying this technique to the three published phase curves, we obtain nightside brightness temperatures of $1173\pm12~$K, $697 \pm 55~$K, and $706\pm26~$K, for the WFC3 and \emph{Spitzer} $3.6$ and $4.5~\mu$m observations, respectively.  The uncertainties were estimated using a $10^{5}$ step Monte Carlo.

We compute the error weighted mean of the brightness temperatures to estimate \added{an average} nightside temperature and propagate uncertainties via Monte Carlo. We obtain a value of $T_{\rm n} = 1076 \pm 11~$K, much higher than the previous value $254\pm182~$K, estimated by \citet{2017arXiv170705790S}, who also used the error weighted mean and propagated uncertainties via Monte Carlo. \added{The new nightside temperature is significantly higher than the previous value ($>4\sigma$ discrepant).}

\added{If our updated brightness temperatures are taken at face value, then the nightside of WASP-43b bears a striking resemblance to the predicted emission spectrum of an isolated brown dwarf with an effective temperature of 600~K \citep{Morley2012}. It must be noted, however, that while setting certain longitudes on a planet's brightness map to zero is better than having negative values, it is still unrealistic. Even neglecting irradiation, hot Jupiters are predicted to have a remnant heat of formation of 50--75~K \citep{2006ApJ...640.1063B}. Our nightside brightness temperatures and effective temperature are probably best thought of as lower limits.}

\section{Discussion \& Conclusions}
Using our dayside and nightside temperature estimates, we can infer the planet's Bond albedo, $A_{\rm B}$,  and heat recirculation efficiency, $ \varepsilon $, using the equations from \citet{2011ApJ...729...54C},
\begin{subequations}
\begin{equation}
T_{\rm d}= T_{0}(1-A_{B})^{1/4}\left(\frac{2}{3}-\frac{5}{12}\varepsilon \right)^{1/4}, 
\end{equation}
\begin{equation}
T_{\rm n}= T_{0}(1-A_{B})^{1/4}\left(\frac{\varepsilon}{4} \right)^{1/4},
\end{equation}\end{subequations} 
where $T_{0} \equiv T_{\star}\sqrt{R_{\star}/a}$ is the planet's irradiation temperature. Both $A_{\rm B}$ and $\varepsilon$ range from 0 to 1.  

\begin{figure}[htb]
\epsscale{1.2}
\plotone{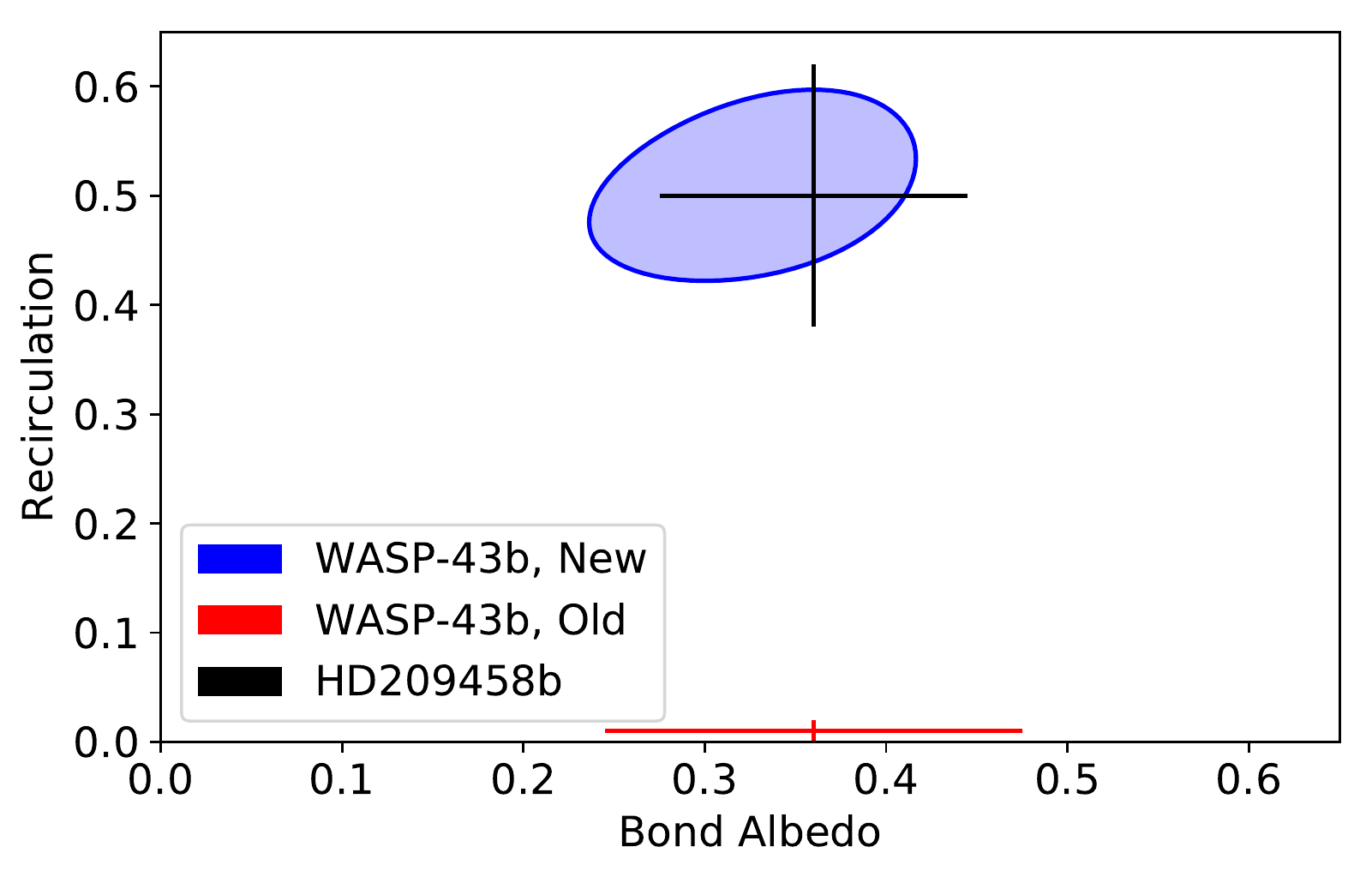}
\caption{Energy budget of WASP-43b. The blue region is the $1\sigma$ contour after our reanalysis. We also plot the best fit values and uncertainties for WASP-43b before correcting the brightness maps and for HD~209458b, both from \citet{2017arXiv170705790S}. WASP-43b no longer hugs the bottom of the plot after our reanalysis, but is now similar to HD~209458b, which we would expect given the similar irradiation temperatures of the two planets \citep{2011ApJ...729...54C,2013ApJ...776..134P,2015MNRAS.449.4192S, Komacek2016}. \label{chi2surf}}
\end{figure}

We use a $10^{5}$ step MCMC to propagate uncertainties and find $A_{\rm B} = 0.3 \pm 0.1$ and $\varepsilon = 0.51 \pm 0.08$. We plot the 1$\sigma$ contour in the $A_{\rm B}$--$\varepsilon$ plane in Figure~\ref{chi2surf}. The Bond albedo and heat recirculation efficiency were previously found to be $A_B =0.36^{+0.11}_{-0.12}$  and $\varepsilon =0.01^{+0.01}_{-0.01}$ \citep{2017arXiv170705790S}, while \cite{2017AJ....153...68S} reported $A_B = 0.19^{+0.08}_{-0.09}$ and $\varepsilon = 0.002^{+0.01}_{-0.002}$.\footnote{We convert their reported $\mathcal{F}$ to $\varepsilon$ using algebra.} Our revised estimate of the Bond albedo is consistent with previous estimates, and indeed with the NIR geometric albedo we inferred above \added{(this may be a coincidence, as only a small fraction of the incident stellar flux is in the WFC3 band).} Our heat transport efficiency, on the other hand, is much greater than previously reported.    

By demanding physically possible brightness maps, our estimate of the planet's energy budget has changed dramatically. Our updated energy budget puts WASP-43b in the same part of $A_{\rm B}$--$\varepsilon$ parameter space as HD~209458b,  and in line with the trend that planets with lower irradiation temperatures have higher heat recirculation (WASP-43b and HD209485b have similar irradiation temperatures). In other words, WASP-43b is no longer an outlier with inexplicably low day\--night heat transport. The models of \cite{2015ApJ...801...86K} may not be missing crucial physics after all. 

Doctoring the brightness maps of WASP-43b is the best one can do without completely refitting the phase curves. For best results, the condition of non-negative brightness maps should be used as a constraint when fitting phase curve parameters simultaneously with astrophysical and detector noise sources. 

Additionally, we have found that reflected light matters in the near infrared for WASP-43b, and by extension for other hot Jupiters. Previous estimates of the dayside temperature of WASP-43b were probably too high, because reflected light may make up a significant portion of the light measured in the WFC3 1.1--1.7$~\mu m$ bandpass. Reflected light has been neglected for all other planets with WFC3 dayside emission spectra, including TrES-3b, WASP-4b, WASP-12b, WASP-33b, WASP-103b CoRoT-2b, HD 189733b, and HD 209458b --- these planets may also exhibit reflected light in the near infrared, and merit a second look.

\acknowledgments
The authors acknowledge support from the McGill Space Institute and l'Institut de recherche sur les exoplan\`etes.  D. K. is supported by a Technologies for Exo-Planetary Science fellowship, and a Centre de recherche en astrophysique du Qu\'ebec fellowship. Thanks to Aisha Iyer for providing the representative water band transmission spectrum, and to Joel Schwartz for providing a table of published planetary phase curve parameters. Thanks to James Xu for helping with the effective temperature estimate. Thanks to Jacob Bean, Taylor Bell, Laura Kreidberg, and Caroline Morley for helpful feedback on the manuscript.

\end{document}